\documentclass[aps, pra, reprint, amsmath, amssymb, groupedaddress, acknowledgments]{revtex4-1}

\usepackage{graphicx, subfigure, dcolumn}
\usepackage{bm,xcolor}

\begin{document}

\title{Cavity-free quantum optomechanical cooling by atom-modulated radiation}

\author{Hoi-Kwan Lau} 
\email[Email address:]{ hklau.physics@gmail.com}
\thanks{\\Current address: Institute of Molecular Engineering, The University of Chicago, 5640 South Ellis Avenue, Chicago, IL 60637, U.S.A.}
\author{Alexander Eisfeld}
\author{Jan-Michael Rost}
\affiliation{Max Planck Institute for the Physics of Complex Systems, N\"{o}thnitzer Stra{\ss}e 38, 01187 Dresden, Germany}

\date{\today} 

\begin{abstract}
We theoretically study the radiation-induced interaction between the mechanical motion of an oscillating mirror and a remotely trapped atomic cloud.  When illuminated by continuous-wave radiation, the mirror motion will induce red and blue sideband radiation, which respectively increases and reduces motional excitation.  We find that by suitably driving $\Lambda$-level atoms, the mirror correlation with a specific radiation sideband could be converted from the outgoing to the incoming radiation.  Such process allows us to manipulate the heating and cooling effects.  Particularly, we develop an optomechanical cooling strategy that can mutually cancel the heating effect of the outgoing and incoming radiations, thus the motional ground state is attainable by net cooling.  Without the necessity of cavity installation or perfect alignment, our proposal complements other efforts in quantum cooling of macroscopic objects.
\end{abstract}


\maketitle

\section{Introduction}

The advancement of science and technology is strongly driven by increasing the precision of mechanical devices.  According to our current understanding of physics, the ultimate precision of mechanical motion is imposed by quantum fluctuation.  Reducing motion of a macroscopic object to the quantum limit allows us to build devices with unprecedented precision for detecting gravitational waves \cite{1980PhRvL..45...75C, 1996PhLA..218..164B, 1997PhLA..232..340B, 1997PhLA..233..303R}, testing fundamental physics \cite{Pikovski:2012hp,Marshall:2003kj}, quantum information processing \cite{Rabl:2010kk, 2013PhRvL.110l0503R}, and more \cite{Poot:2012fh, 2012PhT....65g..29A}.  In practice, the motional fluctuation of most devices is orders of magnitude higher than the quantum limit, due to the inevitable coupling to the environment that induces thermal noise.  

For over a century, great effort has been spent to tackle thermal noise through advancing cooling techniques \cite{book:cool}.  Among which, a promising approach is optomechanical cooling, which dissipates motional excitation by converting it to electromagnetic radiation \cite{WilsonRae:2007jp, Marquardt:2007dn,2014RvMP...86.1391A}.  
Efficient optomechanical cooling usually requires the mechanical oscillator to be placed in an optical cavity, in order to increase the photon-phonon interaction time.  Recently, orders of magnitude reduction of motional excitation has been demonstrated \cite{2011PhRvA..83f3835R,  2015PhRvL.114d3601S}, and a final excitation at single phonon level has been achieved \cite{Teufel:2011jg, Chan:2011dy}.

Nevertheless, the technical challenges to combine both a high-quality mechanical oscillator and a high-quality radiation cavity compromise the applicability of cavity optomechanical cooling.  If the cavity is bad or even unavailable, achieving the motional ground state requires enhancing the cooling efficiency by additional mechanism.  Thanks to the versatile techniques developed in atomic control via electromagnetic radiation, coupling a mechanical oscillator to trapped atoms emerges as a promising option.  Early proposals suggest a standing wave configuration such that the light beams incoming to and reflected from a mirror are aligned to form an optical lattice atomic trap \cite{2010PhRvA..82b1803H, 2011PhRvL.107v3001C, 2013PhRvA..87b3816V, 2015NJPh...17d3044V, Vochezer:2018cq}.  An effective coupling between the mirror motion and the atomic motional or internal state can be established through photon exchange to remove motional excitation sympathetically by laser cooling of atoms.  Recently, some of us have proposed mirror-atom coupling with the incoming and reflected radiation are misaligned and distinct in frequency \cite{2016PhRvA..93b3816S}.  By incorporating electromagnetic induced transparency (EIT), the atoms modulate the sideband radiation that is induced by the mirror motion.  This effect can be used to amplify or damp the classical oscillation of the mirror.

In this paper, we extend the idea in Ref.~\cite{2016PhRvA..93b3816S} to implement optomechanical cooling in the quantum regime.  We consider a setup with an oscillating mirror illuminated by two radiation beams.  
The mechanical motion will generate red and blue sidebands in the reflected radiation.  Note that each sideband contributes oppositely to motional excitation: the blue sideband is created by beam-splitting that cools the mirror, and the red sideband is excited by two-mode-squeezing that increases motional excitation.

Remote from the mirror, we consider a cloud of $\Lambda$-level atom which is trapped at the intersection of an incoming and an outgoing radiation beam. 
By driving the atoms appropriately, we find that the mirror correlation with a specific sideband can be converted from the outgoing to the incoming radiation.  This allows us to develop two cooling strategies: converting the blue sideband to enhance the cooling effect, or converting the red sideband to suppress the heating effect.  We show that the later strategy is particularly promising because the mirror can be cooled to the ground motional state.

Our paper is organised as follow.  The mirror-radiation interaction is first discussed in Sec.~\ref{sec:light}.  The dynamics of the trapped atoms is then studied in Sec.~\ref{sec:atom}.  In Sec.~\ref{sec:cascaded}, we connect the atom and mirror dynamics through a time-local dynamic equation for any system operator.   
In Sec.~\ref{sec:cooling}, we discuss two cooling strategies, and analyse their performance through solving the dynamic equation of motional excitation.  In Sec.~\ref{sec:implement}, we briefly discuss the applicability of our scheme to cool realistic oscillators.  Our paper is concluded in Sec.~\ref{sec:conclusion}.

\section{Light-mirror interaction \label{sec:light}}

\begin{figure}
\begin{center}
\includegraphics[width=\linewidth]{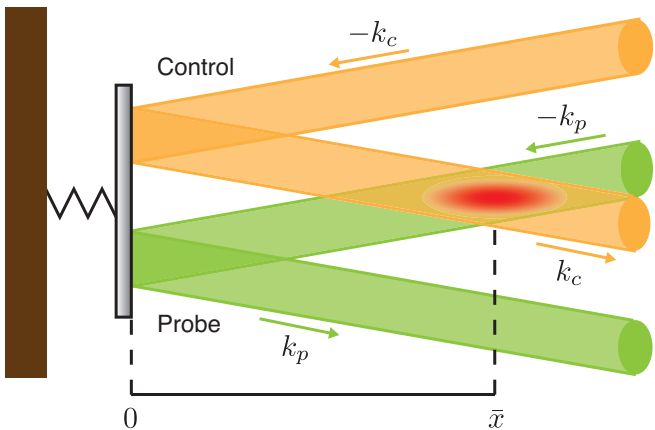}
\caption{ \label{fig:setup} Configuration of our system.  Two radiation beams, Control (orange) and Probe (green), are applied to and reflected from an oscillating mirror.  A diluted atomic cloud (red oval) is trapped at the overlap of the outgoing Control and incoming Probe beams.  }
\end{center}
\end{figure}

The setup of our system is schematically shown in Fig.~\ref{fig:setup}.  An oscillating mirror is illuminated by two beams of continuous-wave (CW) radiation, labelled with Control and Probe, according to the convention of EIT.  The Control beam is applied directly onto the mirror.  Its reflected beam will then be directed through a remotely trapped atomic cloud (distance $\bar{x}$ from the mirror).  On the other hand, the Probe beam passes through the atoms before hitting the mirror.  Its reflected beam will not be collected but dissipated to the environment.  Both the applied Probe and Control radiations are monochromatic classical drives.  Their frequencies are respectively $\omega_{p0}$ and $\omega_{c0}$, and their amplitudes are $\tilde{\alpha}_p$ and $\tilde{\alpha}_c$, which are related to the radiation power as $\mathcal{P}_{p}=\frac{|\tilde{\alpha}_p|^2 c \hbar \omega_{p0} }{2 \pi}$ and $\mathcal{P}_{c}=\frac{|\tilde{\alpha}_c|^2 c \hbar \omega_{c0} }{2 \pi}$ respectively.  

We consider only the fundamental mode of mirror oscillation, while higher order modes can be similarly added to the analysis.  We model the oscillation by a simple harmonic oscillator with frequency $\nu$ and effective mass $m$.  In most systems of interest, the thermal fluctuation of the mirror position is much shorter than the wavelength of the radiation, so that the incoming and outgoing radiation are dominated by the classical drive.  The quantum effect can be studied by considering only the leading order of quantum fluctuations.

The Hamiltonian around the mirror surface is given by
\begin{eqnarray}\label{eq:fullH}
H&=& \hbar \nu \hat{b}^\dag \hat{b}+ \int_{k_{p0}-\kappa}^{k_{p0}+\kappa} \hbar \Delta_{p} \big(\hat{a}^\dag_{k_p}\hat{a}_{k_p} + \hat{a}^\dag_{-k_p}\hat{a}_{-k_p} \big) dk_p  \\
&&+ \int_{k_{c0}-\kappa}^{k_{c0}+\kappa} \hbar \Delta_{c} \big( \hat{a}^\dag_{k_c}\hat{a}_{k_c} + \hat{a}^\dag_{-k_c}\hat{a}_{-k_c} \big) dk_c\nonumber \\
&&+ \frac{\hbar}{2 } \left(\mu_p (\hat{a}^\dag_{p}  -\hat{a}^\dag_{-p})(\hat{b}+\hat{b}^\dag)  +\mu_p^\ast (\hat{b}+\hat{b}^\dag) (\hat{a}_{p} -\hat{a}_{-p} )   \right) \nonumber \\
&&- \frac{\hbar}{2}\left( \mu_c  (\hat{a}^\dag_{c} -\hat{a}^\dag_{-c}) (\hat{b}+\hat{b}^\dag) + \mu_c^\ast (\hat{b}+\hat{b}^\dag) (\hat{a}_{c}  -\hat{a}_{-c} )  \right)~, \nonumber 
\end{eqnarray}
where $k_{p0}\equiv \omega_{p0}/c$ and $k_{c0}\equiv \omega_{c0}/c$.
We have assumed Probe and Control radiation are distinguishable either by sufficiently separated frequencies, or by other degrees of freedom, e.g. polarisation.
We pick a sufficiently wide frequency domain $2c\kappa$ around each classical drive frequency, i.e., $c\kappa \gg \nu$, so that the collection of radiation modes in each domain can be treated as a continuum.  We denote the radiation mode in the continuum around Probe (Control) drive as a \textit{Probe mode} (\textit{Control mode}).   Unless specified, all our integration over wavevector will be assumed conducted over the domain of $2\kappa$.

The first term in Eq.~(\ref{eq:fullH}) is the bare Hamiltonian of mirror motion, where $\hat{b}$ is the annihilation operator of the oscillation mode.  The second and third terms are the bare Hamiltonian of the Probe and Control modes.  The Probe (Control) mode annihilation operator, wavevector, and detuning from classical drive are respectively $\hat{a}_{k_p}$, $k_p$, and $\Delta_p \equiv c|k_p| - \omega_{p0} $ ($\hat{a}_{k_c}$, $k_c$, and $\Delta_c \equiv c|k_c| - \omega_{c0} $).    
In our setup, the incoming and outgoing modes are almost perpendicular to the mirror surface, but not collinear due to misalignment.  The wavevector of each mode is represented by a positive scalar, $k_p$ or $k_c$, and a sign $+$ ($-$) to indicate the outgoing (incoming) propagation.

The fourth and fifth terms in Eq.~(\ref{eq:fullH}) are the leading order optomechanical interaction between the radiation and mirror motion.  This interaction originates from the change of radiation energy density due to mirror motion.  Details of the derivation can be found in Appendix \ref{app:H_mirror}.  The interaction strength is characterised by the factors 
\begin{equation}\label{eq:def_mu}
\mu_p \equiv 2 \sqrt{\frac{c}{2 \pi}} k_{p0} q_0 \tilde{\alpha}_p~~\textrm{and}~~
\mu_c \equiv 2 \sqrt{\frac{c}{2 \pi}} k_{c0} q_0 \tilde{\alpha}_c,
\end{equation}
where the quantum fluctuation of the mirror position is $q_0\equiv \sqrt{\frac{\hbar}{2m\nu}}$.  
For clarity, we denote an annihilation operator with subscript $k$ as a \textit{mode} operator, while that without as a \textit{field} operator, which is defined by the transformation
\begin{equation}
\hat{a}_{\pm p} \equiv \sqrt{\frac{c}{2 \pi}} \int \hat{a}_{\pm k_p} dk_p~~\textrm{and}~~\hat{a}_{\pm c} \equiv \sqrt{\frac{c}{2 \pi}} \int \hat{a}_{\pm k_c} dk_c. \label{eq:field_operator}
\end{equation}

The optomechanical interaction is usually weak in free-space systems, so the dominant interaction is that on resonance.
By using the definition Eq.~(\ref{eq:field_operator}), we observe two types of resonant optomechanical interaction in Eq.~(\ref{eq:fullH}).  The first type couples the blue sideband modes with the mirror motion in the form of $\hat{a}_{k}^\dag \hat{b} +$ h.c., where the blue sideband wavevector can be $k= \pm(k_{p0}+\nu/c)$ or $\pm(k_{c0}+\nu/c)$.  This type of interaction is known as beam-splitting (BS), which converts excitation from one mode to another.  If the input blue sideband mode is in the vacuum, motional excitation will be converted to photons.  The mirror is then cooled if the blue sideband mode is not backcoupled.

The second type couples the red sideband mode with the mirror motion in the form of $\hat{a}_{k} \hat{b} +$ h.c., where the red sideband wavevector is $k=\pm(k_{p0}-\nu/c)$ or $\pm(k_{c0}-\nu/c)$.  This type of interaction is known as two-mode-squeezing (TMS) \cite{Weedbrook:2012fe}.  If the input radiation is in the vacuum, TMS will create both motional excitation and red sideband photons.  The mirror is then heated if the red sideband mode is not backcoupled.

To study the dynamics of motional excitation, we apply the input-output formalism \cite{book:Gardiner_Zoller} to derive the quantum Langevin equation for any mirror operator $\hat{O}_b$,
\begin{eqnarray}
\dot{\hat{O}}_b &=&\mathcal{L}_b (\hat{O}_b) \nonumber \\
&\equiv& i \nu [\hat{b}^\dag \hat{b}, \hat{O}_b] + \frac{|\mu_p|^2+|\mu_c|^2}{2} \mathcal{D}[\hat{b}+\hat{b}^\dag](\hat{O}_b)\nonumber \\
&&- i\mu_p \hat{a}^{(1)\dag}_{-p}[\hat{b}+\hat{b}^\dag,\hat{O}_b]  -i \mu_p^\ast [\hat{b}+\hat{b}^\dag,\hat{O}_b] \hat{a}^{(1)}_{-p} \nonumber \\
&&+ i \mu_c \hat{a}^{\textrm{in}\dag}_{-c}  [\hat{b}+\hat{b}^\dag,\hat{O}_b]  +i \mu_c^\ast [\hat{b}+\hat{b}^\dag,\hat{O}_b] \hat{a}^{\textrm{in}}_{-c}  ~. \label{eq:me_Ob}
\end{eqnarray}
$\hat{O}_b$ can be any polynomial of $\hat{b}$ and $\hat{b}^\dag$.  $\hat{a}^{(1)\dag}_{-p}$ and $\hat{a}^{\textrm{in}\dag}_{-c}$ are respectively the input operator of Probe and Control field \cite{book:Gardiner_Zoller}.  The dissipator super-operator is defined as $\mathcal{D}[\hat{d}]\hat{O}\equiv \hat{d}^\dag \hat{O}\hat{d} -\frac{1}{2}\hat{d}^\dag\hat{d}\hat{O}-\frac{1}{2}\hat{O}\hat{d}^\dag\hat{d}$.  The incoming Control field is assumed to be vacuum, while the incoming Probe field contains information from the trapped atoms.

\section{Light-atom interaction \label{sec:atom}}

\begin{figure}
\begin{center}
\includegraphics{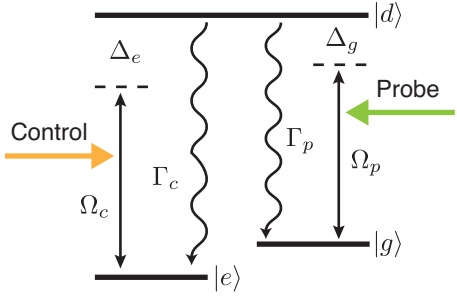}
\caption{ \label{fig:Lambda} Level diagram of a $\Lambda$-level atom.}
\end{center}
\end{figure}

We consider a cloud of atom trapped remotely from the mirror.  For each atom, we utilise only two metastable states, $|g\rangle$ and $|e\rangle$, and one quickly decaying state $|d\rangle$.  The atomic states are arranged in a $\Lambda$-configuration, as shown in Fig.~\ref{fig:Lambda}.  The $|g\rangle \leftrightarrow |d\rangle$ and $|e\rangle \leftrightarrow |d\rangle$ transitions are respectively driven by the incoming Probe and outgoing Control drives.  The total Hamiltonian of the atomic cloud and the radiation is 
\begin{equation}
H_\textrm{total} = H_{p} + \sum_{i}^{N_p} (H_{a}^{(i)} + H_{b}^{(i)} + H_{I}^{(i)} + H_{Ib}^{(i)})~,
\end{equation}
where $N_p$ is the total number of atoms; the index of an atom is arranged according to its distance from the mirror. $H_{p} = \int \hbar \Delta_p \hat{a}^\dag_{-k_p}\hat{a}_{-k_p} dk_p + \int \hbar \Delta_c \hat{a}^\dag_{k_c}\hat{a}_{k_c} dk_c$ is the bare Hamiltonian of the incoming Probe and outgoing Control modes.  $H_{a}^{(i)}$ is the Hamiltonian for the $i$th atom,
\begin{eqnarray}
H_a^{(i)} &=& -\hbar \Delta_g \sigma_{gg}^{(i)} -\hbar \Delta_e \sigma_{ee}^{(i)} + i \hbar\frac{\Omega_p^{(i)\ast}}{2} \sigma_{gd}^{(i)} -  i \hbar\frac{\Omega_p^{(i)}}{2} \sigma_{dg}^{(i)} \nonumber \\
&&+ i \hbar\frac{\Omega_c^{(i)\ast}}{2} \sigma_{ed}^{(i)} -  i \hbar\frac{\Omega_c^{(i)}}{2} \sigma_{de}^{(i)} ~,\label{eq:Ha}
\end{eqnarray}
where $\Delta_g$ ($\Delta_e$) is the detuning of the $|g\rangle \leftrightarrow |d\rangle$ ($|e\rangle \leftrightarrow |d\rangle$) transition from the Probe (Control) drive frequencies.  The Rabi frequency of each atom is the same in magnitude, but different in a position-dependent phase, i.e., $\Omega_p^{(i)} \equiv e^{-i \phi_{pi}} \Omega_p\equiv e^{-i \phi_{pi}} \sqrt{\frac{2c\gamma_p}{\pi}} \tilde{\alpha}_p$, and $\Omega_c^{(i)} \equiv e^{i \phi_{ci}} \Omega_c \equiv e^{i \phi_{ci}} \sqrt{\frac{2c\gamma_c}{\pi}} \tilde{\alpha}_c$, where $\phi_{pi}\equiv \omega_{p0}\frac{x_i}{c}$, $\phi_{ci}\equiv \omega_{c0}\frac{x_i}{c}$, and $x_i$ is the position of the $i$-th atom.  The atomic coherence operator is $\sigma_{ll'}^{(i)}\equiv |l\rangle^{(i)} \langle l'|^{(i)}$, where $|l\rangle^{(i)}$ is the $|l\rangle$ state of the $i$th atom.  

$H_{I}^{(i)}$ is the interaction between the atom and Probe and Control modes, 
\begin{eqnarray}
H_{I}^{(i)} &=& i \hbar \sqrt{\frac{c \gamma_p}{2\pi}} \int e^{ik_p x_i}\hat{a}^\dag_{-k_p} \sigma_{gd}^{(i)} - e^{-ik_p x_i}\sigma_{dg}^{(i)}\hat{a}_{-k_p} dk_p \nonumber \\
&&+ i \hbar\sqrt{\frac{c \gamma_c}{2\pi}} \int e^{-ik_c x_i} \hat{a}^\dag_{k_c} \sigma_{ed}^{(i)} - e^{ik_c x_i} \sigma_{de}^{(i)}\hat{a}_{k_c} dk_c~. \nonumber 
\end{eqnarray}
$\gamma_p$ and $\gamma_c$ are the decay rate to the Probe and Control modes respectively.  $H_{b}^{(i)}$ is the bare Hamiltonian of the bath modes; $H_{Ib}^{(i)}$ is the atom-bath interaction that induces atomic decay.  Here we assumed the atomic cloud is diluted, so each atom is coupled to independent baths.

After integrating the Heisenberg equation for the field operators, and applying standard approximations, we derive the Langevin equation for any atomic operator $\sigma^{(i)}$,
\begin{eqnarray} \label{eq:le_sigmai} 
\dot{\sigma}^{(i)} &=& \mathcal{L}_a^{(i)}(\sigma^{(i)}) \nonumber \\
&\equiv& \frac{i}{\hbar} [H_a^{(i)}, \sigma^{(i)}]  \nonumber \\
&&+ (\gamma_p+\Gamma_p) \mathcal{D}[\sigma_{gd}^{(i)}] (\sigma^{(i)}) + (\gamma_c+\Gamma_c) \mathcal{D}[\sigma_{ed}^{(i)}] (\sigma^{(i)})  \nonumber \\
&&- \sqrt{\gamma_p}e^{i\phi_{pi}} \hat{a}^{(i+1)\dag}_{-p}(t+\frac{x_i}{c}) [\sigma_{gd}^{(i)},\sigma^{(i)}] \nonumber \\
&&+ \sqrt{\gamma_p} e^{-i\phi_{pi}}[\sigma_{dg}^{(i)},\sigma^{(i)}] \hat{a}^{(i+1)}_{-p}(t+\frac{x_i}{c}) \nonumber \\
&&- \sqrt{\gamma_c} e^{-i\phi_{ci}}\hat{a}^{(i)\dag}_{c}(t - \frac{x_i}{c}) [\sigma_{ed}^{(i)},\sigma^{(i)}] \nonumber \\
&&+ \sqrt{\gamma_c} e^{i\phi_{ci}} [\sigma_{de}^{(i)},\sigma^{(i)}] \hat{a}^{(i)}_{c}(t - \frac{x_i}{c})  \nonumber \\
 &&- \sqrt{\Gamma_p} (\hat{r}^{\textrm{in}(i)\dag}_{p} [\sigma_{gd}^{(i)},\sigma^{(i)}] - [\sigma_{dg}^{(i)},\sigma^{(i)}]\hat{r}^{\textrm{in}(i)}_{p})  \nonumber \\
&& - \sqrt{\Gamma_c} (\hat{r}^{\textrm{in}(i)\dag}_{c} [\sigma_{ed}^{(i)},\sigma^{(i)}] - [\sigma_{de}^{(i)},\sigma^{(i)}]\hat{r}^{\textrm{in}(i)}_{c})  ~.
\end{eqnarray}
If not explicitly indicated, the operators are evaluated at time $t$.

\begin{figure}
\begin{center}
\includegraphics{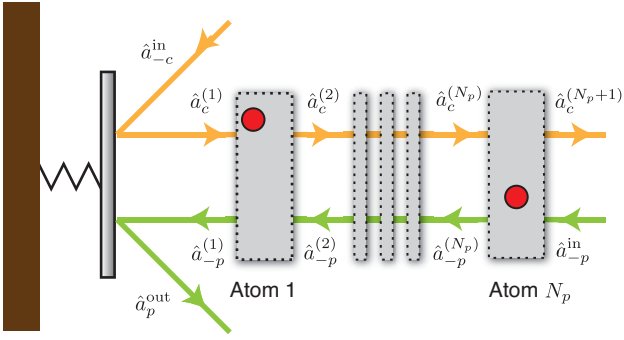}
\caption{ \label{fig:arrange}  The atomic cloud is modelled as an array of atom that the radiation passes through each in sequence.  For a general atomic cloud, we can divide the cloud into parallel slices (grey rectangles), and each slice consists of a negligible number of atoms. For simplicity, in this work we consider each slice contains exactly one atom, and the slices are indexed according to the distance from the mirror.   Our model remains valid if the atoms in each slice are weakly interacting, which is a usual assumption for diluted cloud.
}
\end{center}
\end{figure}

$\Gamma_p$ and $\Gamma_c$ are the spontaneous decay rate from $|d\rangle$ to $|g\rangle$ and $|e\rangle$ respectively, and $\hat{r}_p^{\textrm{in}(i)}$ and $\hat{r}_c^{\textrm{in}(i)}$ are the input field operators of the bath that is responsible for the respective decay.  
$\hat{a}_{-p}^{(i)}$ and $\hat{a}_{c}^{(i)}$ are the Probe and Control input field operators between the $i$-th and $(i-1)$-th atom, which are defined as
\begin{equation}\label{eq:a_i}
\hat{a}_{-p}^{(i)} \equiv \hat{a}_{-p}^\textrm{in} + \mathcal{A}_p^{(i)}~~\textrm{and}~~\hat{a}_c^{(i)} \equiv \hat{a}_c^{(1)}+\mathcal{A}_c^{(i)}~,
\end{equation}
where the collective atomic operators are
\begin{eqnarray}\label{eq:def_Ap}
\mathcal{A}_p^{(i)}(t) &\equiv& \sqrt{\gamma_p} \sum_{j= i}^{N_p} e^{i \phi_{pi}}\sigma^{(j)}_{gd} (t-\frac{x_j}{c}) \\
\mathcal{A}_c^{(i)}(t) &\equiv& \sqrt{\gamma_c} \sum_{j=1}^{i-1} e^{-i \phi_{ci}} \sigma^{(j)}_{ed}(t+\frac{x_j}{c})~.\label{eq:def_Ac}
\end{eqnarray} 
The arrangement of the atoms and field operators are shown in Fig.~\ref{fig:arrange}.

For each atom, the influence from the classical drive is much stronger than that from the sidebands and other atoms.  Therefore the equilibrium state of each atom is well approximated by its bare steady state.
When $\Delta_g=\Delta_e\equiv \Delta$, the bare steady state of a $\Lambda$-level atom is a dark state, 
\begin{equation}
|\textrm{DS}\rangle^{(i)} = \frac{\Omega_c^{(i)}}{\sqrt{|\Omega_p|^2 + |\Omega_c|^2}} |g\rangle^{(i)} - \frac{\Omega_p^{(i)}}{\sqrt{|\Omega_p|^2 + |\Omega_c|^2}} |e\rangle^{(i)}~.
\end{equation}
This state is dark because the radiation transition component vanishes, i.e., $\langle \textrm{DS}|^{(i)}\sigma_\textrm{gd}^{(i)}|\textrm{DS}\rangle^{(i)}=\langle \textrm{DS}|^{(i)} \sigma_\textrm{ed}^{(i)}|\textrm{DS}\rangle^{(i)} = 0$.

As we will discuss, our cooling scheme requires the manipulation of mirror motion correlation with the radiation sidebands.  If the atomic bare steady state is not dark, significant portion of Probe and Control radiation will be scattered to the bath and lost.  Such loss will reduce cooling efficiency.  Therefore, our studies focus on the choice of atomic parameters that the bare steady state is dark.  When the sideband and other atoms are considered, their influence on the atom can be treated as a perturbation on the bare steady state.

\section{Radiation-mediated atom-mirror interaction \label{sec:cascaded}}

In our setup, the reflected Control field will contain correlation with the mirror motion.  It is then fed into and interact with the atomic cloud, which in-turn modulates the Probe field.  Subsequently, the Probe field carries the correlation obtained from the atom and shines onto the mirror.
Overall, the mirror and the atomic cloud form a cascaded quantum system that is connected by radiation.   

The dynamics of such cascaded quantum system can be studied by imposing the following input-output relations as the boundary conditions of Eqs.~(\ref{eq:me_Ob}) and (\ref{eq:le_sigmai})  \cite{book:Gardiner_Zoller},
\begin{eqnarray}
\hat{a}^{(1)}_{c}(t) &=& -\hat{a}^{\textrm{in}}_{-c}(t) + i \frac{1}{2}\mu_c \left( \hat{b}(t)+\hat{b}^\dag (t) \right)  \label{eq:io_ct} \\
\hat{a}^{(1)}_{-p}(t) &=& \hat{a}_{-p}^{\textrm{in}}(t) + \mathcal{A}_p^{(1)}(t)~. \label{eq:io_pt}
\end{eqnarray}
Then any operator composing of mirror and atomic operators, e.g., $\hat{O}\equiv \hat{O}_b \otimes \sigma^{(i)}$, follows the combined master equation $\dot{\hat{O}}=\mathcal{L}_b(\hat{O})+\mathcal{L}^{(i)}_a(\hat{O})$ \cite{book:Gardiner_Zoller}.  

Due to the different time-dependence in the Heisenberg operators, e.g. the field operator in the fourth line of Eq.~(\ref{eq:le_sigmai}), the combined master equation is technically difficult to solve.  Such a time difference appears because radiation takes finite time to travel between atoms and mirror.  Nevertheless, in our regime of interest the time dependence can be made local by the following procedures and approximations.

First, the Probe field that carries information of atom $i$ from time $t-x_i/c$ interacts with the mirror at time $t$.  It is natural to relate these properties by rewriting Eq.~(\ref{eq:le_sigmai}) in terms of the advanced atomic operator $\tilde{\sigma}^{(i)}(t)\equiv \sigma^{(i)}(t-x_i/c)$, which is local at time $t$ \cite{book:Gardiner_Zoller}.  

Second, the Control field which interacts with atom $i$ at time $t- x_i/c$ carries mirror information from time $t-2 x_i/c$.  In combination with the effect of the Probe field, this interaction effectively correlates the mirror properties at time $t-2 x_i/c$ with that at time $t$.  Here we recognise that the mirror motion is dominated by its bare Hamiltonian, so the time discrepancy can be solved by approximating the mirror operator as $\hat{b}(t-2 x_i/c) \approx e^{i\nu 2 x_i/c}\hat{b}(t)$.

Third, we consider a sufficiently small atomic cloud that the light travelling time within the cloud is much shorter than the atomic and mirror time scale, i.e., $\frac{|x_{N_p}-x_1|}{c} \ll 1/\nu, 1/\Gamma, 1/\Omega$.  This allows all atoms to approximately share the same time dependence, i.e., $\sigma(t-\frac{x_i-x_j}{c})\approx \sigma(t)$.  We leave the explicit form of the time-local combined master equation to Appendix~\ref{app:combine}.

\section{Cooling strategy \label{sec:cooling}}

By using the time-local combined master equation for $\hat{O}= \hat{b}^\dag \hat{b}$, we obtain the dynamic equation for the mean motional excitation,  
\begin{eqnarray}
\dot{\langle \hat{b}^\dag\hat{b} \rangle}~~ &=&  \frac{|\mu_p|^2+|\mu_c|^2}{2} \nonumber \\
&&- i\mu_p \langle \mathcal{A}_p^{(1)\dag}(\hat{b}-\hat{b}^\dag)\rangle  -i \mu_p^\ast \langle(\hat{b}-\hat{b}^\dag) \mathcal{A}_p^{(1)}\rangle ~.\label{eq:expect_n}
\end{eqnarray}
The first line contains a constant that is always positive, thus always contributes to heating.  This originates from a stronger TMS heating than BS cooling effect in optomechanical interaction.  

The second line, according to Eq.~(\ref{eq:def_Ap}), is related to the atomic properties.  
We recognise that $\langle \hat{b}\mathcal{A}_p^{(1)} \rangle$ and $\langle \hat{b}^\dag \mathcal{A}_p^{(1)} \rangle$ correspond to different types of atom-induced optomechanical interaction.  Because $\hat{b}$ roughly oscillates at $e^{-i\nu t}$, the zero-frequency component of $\langle \hat{b}\mathcal{A}_p^{(1)}\rangle$ is dominated by the $e^{i\nu t}$ component of $\mathcal{A}_p^{(1)}$.  According to Eq.~(\ref{eq:io_pt}), this frequency component contributes to the red sideband Probe mode.  Since the red sideband interacts with mirror through TMS, the zero frequency component of $\langle \hat{b}\mathcal{A}_p^{(1)} \rangle$ can be viewed as an atom-induced TMS interaction.  Similarly, the zero frequency component of $\langle \hat{b}^\dag \mathcal{A}_p^{(1)} \rangle$ corresponds to the atom-induced BS interaction.

To quantify the atomic contributions, we construct the master equation for $\hat{O} = \hat{b}\otimes \sigma^{(i)}$ and $\hat{b}^\dag \otimes \sigma^{(i)}$.  In our regime of weak optomechanical interaction, the expectation value of these operators can be approximated by a Floquet mode expansion in motional frequency, i.e., $\langle \hat{O}(t)\rangle \approx \sum_n \langle \hat{O}\rangle_n (t) e^{i n \nu t}$, where $\langle \hat{O}\rangle_n (t)$ varies in a time scale much slower than $1/\nu$.  Because the bare dynamics of motional excitation is not oscillating, we consider only the dominating atomic effect, i.e., the zero frequency components, $\langle\hat{O}\rangle_0$.
After summing the contributions of each atom, we obtain the recurrence relation
\begin{widetext}
\begin{eqnarray}
\langle \hat{b} \mathcal{A}_p^{(i)} \rangle_0 &=& \Big(1+ |\tilde{\alpha}_c|^2J(-\nu)\Big) \langle \hat{b}\mathcal{A}_p^{(i+1)} \rangle_0 -  \tilde{\alpha}_p\tilde{\alpha}_c^\ast J(-\nu) \langle \hat{b}\mathcal{A}_c^{(i)} \rangle_0 - i e^{-i\nu\tau} \frac{\mu_c}{2} \tilde{\alpha}_p\tilde{\alpha}_c^\ast J(-\nu) \langle\hat{b} \hat{b}^\dag\rangle~, \label{eq:Apb2}\\
\langle \hat{b}\mathcal{A}_c^{(i+1)}\rangle_0 &=& - \tilde{\alpha}_p^\ast \tilde{\alpha}_c J(-\nu) \langle \hat{b}\mathcal{A}_p^{(i+1)} \rangle_0 +\Big(1+ |\tilde{\alpha}_p|^2 J(-\nu) \Big) \langle \hat{b}\mathcal{A}_c^{(i)} \rangle_0 + i e^{-i\nu\tau} \frac{\mu_c}{2} |\tilde{\alpha}_p|^2 J(-\nu) \langle\hat{b}\hat{b}^\dag\rangle ~, \label{eq:Acb2} \\
\langle \hat{b}^\dag \mathcal{A}_p^{(i)} \rangle_0 &=& \Big(1+ |\tilde{\alpha}_c|^2J(\nu)\Big) \langle \hat{b}^\dag\mathcal{A}_p^{(i+1)} \rangle_0 -  \tilde{\alpha}_p\tilde{\alpha}_c^\ast J(\nu) \langle \hat{b}^\dag\mathcal{A}_c^{(i)} \rangle_0 - ie^{i\nu\tau} \frac{\mu_c}{2} \tilde{\alpha}_p\tilde{\alpha}_c^\ast J(\nu) \langle\hat{b}^\dag \hat{b} \rangle~, \label{eq:Apbd2} \\
\langle\hat{b}^\dag\mathcal{A}_c^{(i+1)}\rangle_0 &=& - \tilde{\alpha}_p^\ast \tilde{\alpha}_c J(\nu) \langle \hat{b}^\dag\mathcal{A}_p^{(i+1)} \rangle_0 +\Big(1+ |\tilde{\alpha}_p|^2 J(\nu) \Big) \langle \hat{b}^\dag\mathcal{A}_c^{(i)} \rangle_0 + i e^{i\nu\tau} \frac{\mu_c}{2} |\tilde{\alpha}_p|^2 J(\nu) \langle\hat{b}^\dag\hat{b}\rangle ~, \label{eq:Acbd2}
\end{eqnarray}
\end{widetext}
where $\tau \equiv 2 \bar{x}/c$ is the round-trip traveling time of light between the mirror and the atomic cloud; $\bar{x} = \sum_j x_j/N_p$ is the mean distance of the atoms from the mirror.  The expression of the spectral factor $J(\omega)$ is given by
\begin{widetext}
\begin{equation}
J(\omega) = \frac{\gamma_p \gamma_c}{2\pi} \frac{ i 16 \omega c}{(|\Omega_p|^2+|\Omega_c|^2)(-2 i \omega (\gamma_p + \Gamma_p + \gamma_c + \Gamma_c) +4 \Delta \omega -4 \omega^2 + |\Omega_p|^2+|\Omega_c|^2)}~. \label{eq:J_exp}
\end{equation}
\end{widetext}
We leave the derivation of this recurrence relation to Appendix \ref{app:recurrence}.

Eqs.~(\ref{eq:Apb2})-(\ref{eq:Acb2}) and Eqs.~(\ref{eq:Apbd2})-(\ref{eq:Acbd2}) form two systems of equation that can be solved separately.  For simplicity, we assume that both Probe and Control drives are the same in both phase and amplitude, i.e., $\tilde{\alpha}_p=\tilde{\alpha}_c\equiv\tilde{\alpha}$.
Before presenting the solution, we discuss the physics underneath.  By adding Eq.~(\ref{eq:Apb2}) to (\ref{eq:Acb2}), and (\ref{eq:Apbd2}) to (\ref{eq:Acbd2}), we get the relations
\begin{eqnarray}\label{eq:transfer_red}
\langle \hat{b} \mathcal{A}_p^{(i)} \rangle_0 - \langle \hat{b} \mathcal{A}_p^{(i+1)} \rangle_0 &=& - \big(\langle \hat{b} \mathcal{A}_c^{(i+1)} \rangle_0 - \langle \hat{b} \mathcal{A}_c^{(i)} \rangle_0 \big) \\
\langle \hat{b}^\dag \mathcal{A}_p^{(i)} \rangle_0 - \langle \hat{b}^\dag \mathcal{A}_p^{(i+1)} \rangle_0 &=& - \big(\langle \hat{b}^\dag \mathcal{A}_c^{(i+1)} \rangle_0 - \langle \hat{b}^\dag \mathcal{A}_c^{(i)} \rangle_0 \big)~~~.\label{eq:transfer_blue}
\end{eqnarray}
According to Eqs.~(\ref{eq:a_i}) and (\ref{eq:def_Ap}), and the fact that $\hat{b}$ varies roughly as $e^{i \nu t}$, Eq.~(\ref{eq:transfer_red}) describes the changes of the red sideband modes of Probe and Control after interacting with the $i$th atom.  More explicitly, a reduction of correlations between outgoing Control mode and the mirror will be the same as an increase of correlations between incoming Probe mode and the mirror.  This can be interpreted as a conversion of the correlation with the mirror from the outgoing Control mode to the incoming Probe mode.  Subsequently, the Probe red sideband will bring the correlation back to the mirror and affect the TMS interaction.
Similar phenomenon is observed in Eq.~(\ref{eq:transfer_blue}), where the mirror correlation with the blue sideband mode is converted from outgoing Control to incoming Probe.  This correlation will then affect the BS interaction. 

Following the procedure in Appendix \ref{app:solution}, we obtain the solution for Eqs.~(\ref{eq:Apb2})-(\ref{eq:Acbd2}) as
\begin{eqnarray}\label{eq:bAp}
\langle \hat{b}\mathcal{A}_p^{(1)} \rangle_0 &=& -i e^{-i\nu\tau}\frac{1}{2}\mu_c \frac{N_p |\tilde{\alpha}|^2 J(-\nu)}{1- N_p |\tilde{\alpha}|^2 J(-\nu)} \langle \hat{b} \hat{b}^\dag\rangle \\
\langle \hat{b}^\dag \mathcal{A}_p^{(1)} \rangle_0 &=& -i e^{i\nu\tau} \frac{1}{2}\mu_c \frac{N_p |\tilde{\alpha}|^2 J(\nu)}{1- N_p |\tilde{\alpha}|^2 J(\nu)} \langle \hat{b}^\dag \hat{b}\rangle~. \label{eq:bdAp}
\end{eqnarray}
When substituting Eqs.~(\ref{eq:bAp}) and (\ref{eq:bdAp}) into Eq.~(\ref{eq:expect_n}), we obtain the main result of our work: a dynamic equation for the mean motional excitation,
\begin{equation}\label{eq:main}
\langle \dot{\hat{n}} \rangle = \mathcal{N}_0 +\Lambda_+ \langle \hat{n} \rangle +\Lambda_- \big(\langle \hat{n} \rangle+1 \big)~,
\end{equation}
where the phonon number operator is $\hat{n}\equiv \hat{b}^\dag \hat{b}$, and
\begin{eqnarray}
\mathcal{N}_0 &\equiv&  \frac{|\mu_p|^2+|\mu_c|^2}{2} \\
\Lambda_\pm &\equiv& \pm \textrm{Re}\Big(e^{\pm i\nu\tau} \frac{\mu_p^\ast \mu_c  N_p |\tilde{\alpha}|^2 J(\pm\nu)}{1- N_p |\tilde{\alpha}|^2 J(\pm \nu)} \Big)~.
\end{eqnarray}
$\mathcal{N}_0$ is the heating rate induced by the classical drives, due to the stronger TMS heating than BS cooling.  $\Lambda_+$ comes from atom-induced BS, and $\Lambda_-$ is the contribution of atom-induced TMS.  The atoms induce a net cooling effect on the mirror if $\Lambda_+ + \Lambda_- <0 $.  In this case, the residual motional excitation at the steady state, i.e., $\langle \dot{\hat{n}}\rangle =0$, is 
\begin{equation}
\langle \hat{n}\rangle_\textrm{ss} = -\frac{\mathcal{N}_0 + \Lambda_-}{\Lambda_+ + \Lambda_-}~,
\end{equation}
where the subscript ss denotes the steady state.

The sign of $(\Lambda_+ + \Lambda_- )$ is determined by the phase, $e^{i \nu \tau}$, as well as the relative importance of the atom-induced effects, of which the magnitude is characterised by the spectral factor $J(\omega)$.  
In practice, the phase can be manipulated by choosing the position of the atomic cloud, and the spectral factor can be engineered by adjusting atomic parameters.  

\begin{figure}
\begin{center}
\includegraphics{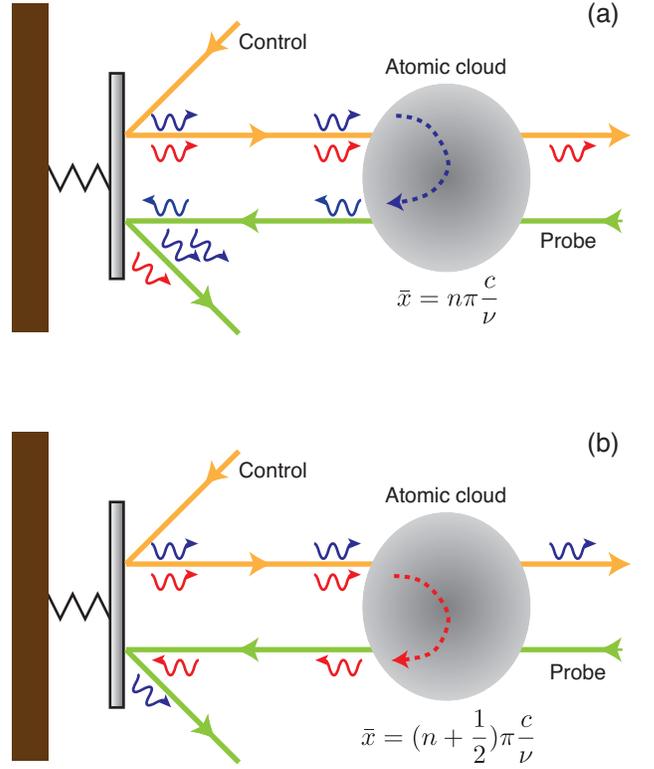}
\caption{ \label{fig:strategy} Illustration of cooling strategies. (a) BS-enhancing strategy: Atomic parameters are chosen to convert blue sideband correlation with mirror from outgoing Control to incoming Probe.  The correlation contained by the Probe mode will enhance the BS cooling effect.  Nevertheless, red sideband of both outgoing Control and incoming Probe will be dissipated and induce TMS heating.  (b) TMS-suppressing strategy: Atomic parameters are chosen to convert red sideband correlation with mirror from outgoing Control to incoming Probe.  Due to a time delay, the correlation contained by the Probe mode will suppress TMS heating.  However, blue sideband of both Control and Probe is unaffected.  Its dissipation will induce net cooling on the mirror.}
\end{center}
\end{figure}

For our purpose of cooling, we explore system parameters that could implement either of the mechanism: enhancing cooling effect of BS, or suppressing heating effect due to TMS.  Illustration of our cooling strategies is shown in Fig.~\ref{fig:strategy}.  In the following, we will show that 
while both mechanisms can induce cooling, the motional ground state cannot be reached by enhancing BS alone.  Ground state cooling requires optimally choosing system parameters to fully suppress the TMS heating effect.

\subsection{Enhancing beam-splitting \label{ssec:BS}}

First, we study the cooling strategy by an atom-enhanced BS interaction.  In this case, the net cooling rate is dominated by a negative $\Lambda_+$.  From Eq.~(\ref{eq:J_exp}), we learn that $J(\omega)$, as well as the proportionality constant $\frac{N_p |\tilde{\alpha}|^2 J(\omega)}{1- N_p |\tilde{\alpha}|^2 J(\omega)}$, always has a negative real part.  Therefore $\Lambda_+$ is always negative if $e^{i\nu \tau}=1$.  Such criterion could be satisfied by placing the atomic cloud close to the mirror, i.e., $\bar{x}\approx 0$, or generally at location $\bar{x}=n\pi\frac{c}{\nu}$.

\begin{figure}
\begin{center}
\includegraphics{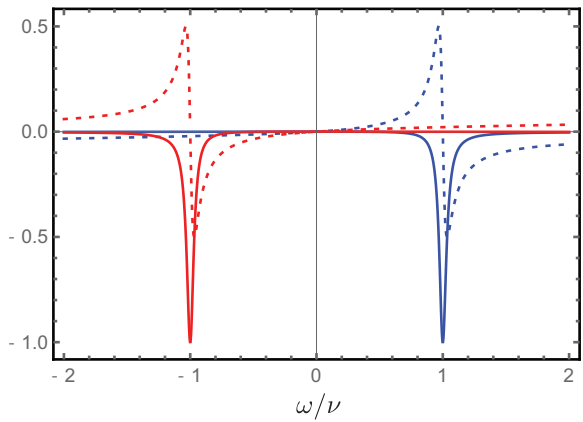}
\caption{ \label{fig:J} Typical behaviour of spectral function $J(\omega)$.  Blue: Real (solid) and imaginary (dotted) part of $J(\omega)$, normalised by $|J(\nu)|$.  The parameters are $\Omega_p = \Omega_c=4 \nu$, $\gamma_p+\Gamma_p = \gamma_c+\Gamma_c = 0.3 \nu$.  The detuning $\Delta$ is chosen to satisfy Eq.~(\ref{eq:blue_detune}), in order to facilitate the atom interaction with blue sideband radiation.  The interaction with red sideband is suppressed, as we can see $|J(\nu)| \gg |J(-\nu)|$.  Red: Real (solid) and imaginary (dotted) part of $J(\omega)$, of which the atomic parameters are the same as those of the blue lines, except $\Delta$ is chosen to satisfy Eq.~(\ref{eq:red_detune}). 
This spectral function is normalised by $|J(-\nu)|$.}
\end{center}
\end{figure}

On the other hand, at this position the atom-induced TMS always contributes to heating, i.e., $\Lambda_->0$.  Achieving a net cooling rate thus requires enhancing the BS effect, i.e., $|\Lambda_+| \gg |\Lambda_-|$.  This can be achieved by introducing a frequency asymmetry in the spectral factor, i.e., $|J(\nu)|\gg |J(-\nu)|$.  Inspired by the parameter choice in EIT cooling \cite{Morigi:2000wv, Morigi:2003cv}, such asymmetry can be produced by using sufficiently large Rabi frequencies, i.e., $|\Omega_p|^2+|\Omega_c|^2 \gg \nu (\gamma_p+\Gamma_p+\gamma_c+\Gamma_c)$, and choosing a detuning that satisfies
\begin{equation}\label{eq:blue_detune}
4\Delta \nu - 4 \nu^2 +  |\Omega_p|^2+|\Omega_c|^2 =0~.
\end{equation}
Typical behaviour of this $J(\omega)$ is shown as the blue lines in Fig.~\ref{fig:J}.

\subsubsection{Few atom regime}

In the regime that the number of atom is small, i.e., $N_p |\tilde{\alpha}|^2 |J(\omega)| \ll 1$ for every $\omega$, the net cooling rate can be approximated as
\begin{equation}\label{eq:Lambda_BS1}
\Lambda_+ + \Lambda_- \approx  N_p |\tilde{\alpha}|^2 \mu_p^\ast \mu_c \textrm{Re}(J(\nu)-J(-\nu))~.
\end{equation}
According to Eq.~(\ref{eq:def_mu}), $\mu_p^\ast \mu_c$ is real and positive under our assumption $\tilde{\alpha}_p = \tilde{\alpha}_c$.  It is easy to show that $\Lambda_+ + \Lambda_- <0$, and so the mirror is cooled in this setup.  

In fact, this is the same setup being studied in Ref.~\cite{2016PhRvA..93b3816S}, which derives the same cooling rate Eq.~(\ref{eq:Lambda_BS1}) by using semi-classical techniques.  Nevertheless, we find that the few-atom assumption implies that the atom-induced cooling is always much weaker than the heating induced by the classical drives, i.e., $\mathcal{N}_0 \gg |\Lambda_+|$.  Unnoticed in Ref.~\cite{2016PhRvA..93b3816S}, the residual motional excitation in this setup is thus enormous, i.e., $\langle \hat{n} \rangle_\textrm{ss} \gg 1$.

\subsubsection{Many atom regime}

In order to improve the cooling rate, we take the number of atom to be sufficiently large to satisfy $N_p |\tilde{\alpha}|^2 |J(\nu)| \gg 1 \gg N_p |\tilde{\alpha}|^2 |J(-\nu)|$.  In this regime, the atom-induced BS effect is saturated as $\Lambda_+ \approx -\mu_p^\ast \mu_c$, while $\Lambda_-\approx -N_p |\tilde{\alpha}|^2 \mu_p^\ast \mu_c \textrm{Re}\big(J(-\nu)\big)$ is small due to an asymmetric  spectral function.  The steady state motional excitation is
\begin{equation}\label{eq:nf_BS}
\langle \hat{n} \rangle_\textrm{ss} \approx \frac{|\mu_p|^2+|\mu_c|^2}{2 \mu_p^\ast \mu_c} \geq 1~,
\end{equation}
where the last relation is imposed by Cauchy-Schwarz inequality.

The above result shows that even with the optimal choice of system parameters, enhancing the BS cooling effect could only cool the mirror to a thermal state with unity motional excitation.  This could be understood by the intuition that, when the mirror is close to motional ground state, BS interaction is ineffective to convert motional excitation to the blue sideband.  On the other hand, TMS effect remains effective to induce excitation even when the red sideband mode is in vacuum.  As a result, the motional ground state would not be a steady state if TMS is not suppressed.

\subsection{Suppressing two-mode-squeezing}

To pursue ground-state cooling, we here consider another cooling strategy that aims to suppress the TMS heating effect.  In this case, the net cooling rate will be dominated by a negative $\Lambda_-$.  This is achievable if the atomic cloud is placed at a distance $\bar{x}=(n+\frac{1}{2})\pi\frac{c}{\nu}$ from the mirror, so that $e^{i\nu\tau}=-1$.  

At this position, the atom-induced BS reduces cooling efficiency, i.e., $\Lambda_+>0$.  A net cooling rate can be attained by engineering an asymmetric spectral factor $|J(-\nu)| \gg |J(\nu)|$, which leads to $|\Lambda_-|\gg |\Lambda_+|$.  Similar to the choice of parameters in Sec.~\ref{ssec:BS}, such asymmetry can be achieved by using large Rabi frequencies, but choosing a detuning that instead satisfies
\begin{equation}\label{eq:red_detune}
-4\Delta \nu - 4 \nu^2 +  |\Omega_p|^2+|\Omega_c|^2 =0~.
\end{equation}
Typical behaviour of this $J(\omega)$ is shown as the red lines in Fig.~\ref{fig:J}.

In the many-atom regime that $N_p |\tilde{\alpha}|^2 |J(-\nu)| \gg 1 \gg N_p |\tilde{\alpha}|^2 |J(\nu)$, we have
$ \Lambda_- \approx -\mu_p^\ast \mu_c$ and $\Lambda_+ \approx -N_p |\tilde{\alpha}|^2 \mu_p^\ast \mu_c \textrm{Re}\big(J(\nu)\big)$.  The atom-induced TMS interaction is saturated, but $|\Lambda_+|$ is small due to the spectral factor asymmetry.  The steady state motional excitation is given by
\begin{equation}
\langle \hat{n} \rangle_\textrm{ss} \approx \frac{|\mu_p|^2 + |\mu_c|^2 - 2 \mu_p^\ast \mu_c}{2 \mu_p^\ast \mu_c} \geq 0~.
\end{equation}

In stark contrast to the BS-enhancing strategy, the TMS-suppressing strategy here could achieve the motional ground steady state if $\mu_p=\mu_c$.  According to Eq.~(\ref{eq:def_mu}), this criterion can be fulfilled by choosing two atomic state transitions with similar energy, so that the classical drive frequency is roughly the same, i.e., $\omega_{p0} \approx \omega_{c0}$  
\footnote{We note that this criterion does not contradict with our assumption that Probe and Control mode frequencies are sufficiently separated.  It is because Probe and Control can be treated as two separate continuum if their frequency difference is much wider than oscillation frequency $\nu$, which is orders of magnitude smaller than the optical frequency of radiation.  Furthermore, Probe and Control can be distinct by other degrees of freedom, such as polarisation.} .

We now explain the principle behind this strategy.  At the beginning, the Control red sideband couples to the mirror motion through TMS interaction.  If this sideband is lost, motional excitation will increase.  Therefore, we choose the system parameters, such that the mirror correlation with the outgoing Control red sideband is completely transferred to the incoming Probe.

The crucial trick of our strategy is that the round-trip travelling time is chosen to satisfy $e^{i\nu \tau}=-1$.  When the Probe field reaches the mirror, the mirror motion is $\pi$-phase behind that at the beginning.  This $\pi$-phase effectively inverts the sign of the TMS Hamiltonian, so the mirror and Probe red sideband undergo an anti-TMS interaction.  If the optomechanical interaction strength is the same for Probe and Control field, i.e., $\mu_p=\mu_c$, the anti-TMS can completely `undo' the mirror-Control TMS.  Therefore, the TMS heating effect is fully suppressed.  

On the other hand, because $|\Lambda_+|$ is kept to be small, the BS cooling effect of the blue sideband is barely affected by the atoms, so the mirror experiences a net cooling.  Without TMS heating, the motional ground state is attainable at the steady state.

\section{Practical implementation \label{sec:implement}}

Finally, we discuss the practicality of our scheme in cooling realistic mechanical oscillator.  We first include environmental heating to the dynamic equation,
\begin{equation}
\langle\dot{\hat{n}} \rangle =  - (|\mu|^2 + \frac{\nu}{Q}) \langle \hat{n}\rangle + \frac{\nu}{Q} \mathcal{N}_\textrm{th}~.
\end{equation}
$Q$ is the quality factor of the oscillator, which is defined as the number of oscillation period to lose half motional excitation when the environment is at zero temperature.  $\mathcal{N}_\textrm{th}\approx \frac{k_B T}{\hbar \nu}$ is the motional excitation at thermal equilibrium.  For simplicity, we have picked the optimal optomechanical strength $\mu_p \approx \mu_c = \mu$, and assumed $\Lambda_+\rightarrow 0$ is suppressed by appropriately chosen system parameters.  

As an example, we consider a state-of-the-art mechanical oscillator reported in Ref.~\cite{2014NatCo...5E3638T}.  This is a single crystal diamond cantilever with length 240~$\mu$m, width 12~$\mu$m, thickness 0.66~$\mu$m, oscillation frequency $\nu \approx 2\pi \times 32$~kHz, and quality factor $Q\approx 1.5 \times10^6$.  

The remotely trapped atoms are chosen to be Rubidium 85.  The atomic transition employed is the D$_2$ line, i.e., $5^2S_{1/2}\leftrightarrow 5^2P_{3/2}$, which couples to the radiation of wavelength $\approx 780$~nm \cite{atom_data}.  The Probe and Control fields can be differentiated by different polarisation of transitions, or by hyperfine state energy shift due to external static magnetic field.  

At a 10 mK environment, which is achievable by state-of-the-art cryogenic techniques \cite{book:cool}, implementing our cooling scheme with CW laser power $\mathcal{P}=10$~mW could reduce the motional excitation from initially $\langle \hat{n} \rangle(t=0)=\mathcal{N}_\textrm{th}\approx 6500$ to a quantum level steady state, $\langle \hat{n}\rangle_\textrm{ss} \approx 2$.  At this level of motional excitation, the oscillator can already implement a variety of application, such as detecting macroscopic non-classicality \cite{2016PhRvA..94a3850M} or quantum computation \cite{Lau:2017et}.  

The steady state motional excitation can be further reduced by, e.g., enhancing the power of CW laser, reducing the environmental temperature, or increasing the zero-point position fluctuation by using a lower frequency oscillator.

\section{Conclusion \label{sec:conclusion}}

In this work, we study the radiation-induced interaction between remotely trapped atoms and the motion of a mirror.  We consider a cavity-free setup that the mirror is driven by two continuous-wave radiations, Control and Probe.  We find that the resonant optomechanical interaction would either be beam-splitting with the blue radiation sideband, or two-mode-squeezing with the red radiation sideband.  We recognise that these interactions contribute respectively to the cooling and heating of the mirror.

Remotely from the mirror, we consider a cloud of atom is trapped at the overlap of outgoing Control and incoming Probe.  We find that both the rate and the type of atom-radiation interaction depend on the atomic parameters.  Particularly, if the $\Lambda$-level atoms are driven in a dark steady state configuration, the mirror correlation with a specific sideband of the outgoing Control can be converted to that of the incoming Probe.  Our main result is provided as a dynamic equation of motional excitation in Eq.~(\ref{eq:main}).

We explore two strategies to utilise this atom-modulated effect to cool the mirror.  By resonantly interacting with the blue sideband, the mirror is cooled due to the enhanced BS interaction.  However, the motional ground state is not attainable by this strategy because of the prevailing TMS heating.  On the other hand, TMS heating effect can be suppressed through resonantly interacting with the red sidebands.  By this strategy, ground state cooling is achievable.

Without the necessity to install high-quality cavity around atoms or mirror, nor to precisely align the incoming and outgoing radiation to form optical lattice, we believe our scheme could broaden the systems that could be optomechanically cooled.  The theoretical tools developed in this work could also be useful to study other radiation-induced interaction between atoms and mirror, such as atom-modulated phonon lasing or atom-mirror entanglement generation.

\begin{acknowledgments}
H.-K.L. thanks MPI-PKS visitor program for its generous support, and Adrian Sanz-Mora and Aashish Clerk for their fruitful discussion.  J.-M.R. and A.E. acknowledge support by the Deutsche Forschungsgemeinschaft (DFG) through Grants No. RO 1157/9-1 and EI 872/4-1 within the Priority Program SPP 1929 (GiRyd).
\end{acknowledgments}

\appendix

\section{Hamiltonian near mirror surface \label{app:H_mirror}}

The total Hamiltonian of the mirror and the radiation field in the vicinity of the mirror surface is given by
\begin{eqnarray}
\frac{H}{\hbar}&=& \nu \hat{b}^\dag \hat{b}+ \int_{-\infty}^\infty  \omega_k \hat{a}^\dag_k\hat{a}_k dk  -\frac{A}{\hbar} \hat{q}\left(  \frac{\epsilon_0}{2}E(0)^2 +  \frac{\mu_0}{2} \mathcal{H}(0)^2 \right) ~. \nonumber\\\label{eq:H_mirror1}
\end{eqnarray}
where $\hat{b}$ is the annihilation operator of the mirror motion; $\hat{a}_k$ is the annihilation operator for the light mode with wave vector $k$; $\omega_k=c|k|$ is the frequency of the mode; $A$ is the cross-section area of the beam; $\hat{q}=q_0(\hat{b}+\hat{b}^\dag)$ is the position operator of the mirror.  The first and second term of Eq.~(\ref{eq:H_mirror1}) are the bare Hamiltonian of the mirror motion and radiation field, respectively.  
The third term is the optomechanical coupling, which is the change of total electromagnetic energy due to the change of radiation space upon mirror displacement.  The electric field and magnetic field operator at position $x$ is defined by
\begin{eqnarray}
E(x) &=& \int^\infty_{-\infty} i \sqrt{\frac{\hbar \omega_k}{4 \pi \epsilon_0 A}} (\hat{a}_k e^{ikx}-\hat{a}^\dag_k e^{-ikx})dk \nonumber \\
\mathcal{H}(x) &=& \int^\infty_{-\infty} \frac{i}{\mu_0} \sqrt{\frac{\hbar}{4 \pi \epsilon_0 A \omega_k}} k(\hat{a}_k e^{ikx}-\hat{a}^\dag_k e^{-ikx})dk~. \label{eq:def_EH} \nonumber
\end{eqnarray}
Because the incoming and outgoing radiations are assumed to be almost along the $x$ direction, 
we can use a scalar $k$ to represent the wavevector of the incoming ($k<0$) and outgoing ($k>0$) radiation.

After extracting the contribution of the classical drive, i.e., $\hat{a}_k\rightarrow \hat{a}_k+\tilde{\alpha}_p e^{-i\omega_{p0}t} (-\delta(k-k_{p0})+\delta(k+k_{p0}))+\tilde{\alpha}_c e^{-i\omega_{c0}t} (\delta(k-k_{c0})-\delta(k+k_{c0}))$, and collecting the quantum contributions up to second order of mode operators, we get the Hamiltonian in Eq.~(\ref{eq:fullH}).  Note that this derivation considers only the radiation of one degree of freedom, and so Probe and Control are distinguished by frequency.  For Probe and Control that are distinct in other degrees of freedom, e.g., polarisation or orbital angular momentum, the Hamiltonian can be derived similarly.

The Hamiltonian in Eq.~(\ref{eq:fullH}) is the starting point of our studies on mirror dynamics.  We first directly integrate the Heisenberg equation of the mode operators, i.e., $\dot{\hat{a}}_k=\frac{i}{\hbar}[H,\hat{a}_k]$.  By using Eq.~(\ref{eq:field_operator}) and the definition Eq.~(\ref{eq:def_mu}), we obtain the relations for the incoming field operators:
\begin{eqnarray} \label{eq:ap_relation}
\hat{a}_{-p}(t) &=& \hat{a}^{(1)}_{-p}(t) + i \frac{1}{4} \mu_p \left( \hat{b}(t)+\hat{b}^\dag (t) \right) \\
\hat{a}_{-c}(t) &=& \hat{a}^{\textrm{in}}_{-c}(t) - i \frac{1}{4} \mu_c \left( \hat{b}(t)+\hat{b}^\dag (t) \right)~,\label{eq:ac_relation}
\end{eqnarray}
where $\hat{a}^{\textrm{in}}_{-c}$ is the Control field input operator from vacuum onto the mirror surface, and $\hat{a}^{(1)}_{-p}(t)$ is the Probe field input operator from vacuum through the atomic cloud onto the mirror surface \cite{book:Gardiner_Zoller}.  
Because the electric field vanishes on the surface of a perfect conductor, the incoming and outgoing field operator obeys the boundary condition
\begin{equation}\label{eq:leftright_relation}
\hat{a}_p + \hat{a}_{-p}=\hat{a}_c + \hat{a}_{-c}=0~.
\end{equation}

Substituting Eqs.~(\ref{eq:ap_relation})-(\ref{eq:leftright_relation}) into the Heisenberg equation of mirror operators, i.e., $\dot{\hat{O}}_b=\frac{i}{\hbar}[H,\hat{O}_b]$, we obtain the Langevin equation in Eq.~(\ref{eq:me_Ob}).

\section{Atomic dynamics and steady state \label{app:atom}}

To study the dynamics of atoms, we arrange the 9 atomic operators as an 8-entry vector, i.e., $\vec{\sigma} \equiv (\sigma_{dd}-\sigma_{gg},\sigma_{ge},\sigma_{eg}, \sigma_{dd}-\sigma_{ee},\sigma_{gd},\sigma_{dg},\sigma_{ed},\sigma_{de} )^\textrm{T}$, where we have extracted the time invariant $\sigma_{gg}+\sigma_{ee}+\sigma_{dd}=1$ \cite{Lau:2016cool}.  The bare dynamics of the atom is described by Eq.~(\ref{eq:le_sigmai})  with no contribution from the mirror and other atoms, i.e., $\hat{a}_{-p}^{(i+1)}$ and  $\hat{a}_{c}^{(i)}$ are taken as vacuum field operators.  The expectation value of atomic operators varies as
\begin{eqnarray}
\langle\dot{\sigma}^{(i)}_{1}\rangle &=& -\Gamma_1 \langle\sigma^{(i)}_{1}\rangle  -\Gamma_1 \langle\sigma^{(i)}_{4}\rangle - \Omega_p^{(i) \ast} \langle \sigma^{(i)}_{gd} \rangle - \Omega_p^{(i)} \langle \sigma^{(i)}_{dg} \rangle \nonumber \\
&&- \frac{\Omega_c^{(i) \ast}}{2} \langle \sigma^{(i)}_{ed} \rangle - \frac{\Omega_c^{(i)}}{2} \langle \sigma^{(i)}_{de} \rangle -\Gamma_1 \nonumber \\
\langle\dot{\sigma}^{(i)}_{ge} \rangle &=& -i (\Delta_g - \Delta_e) \langle \sigma^{(i)}_{ge}\rangle + \frac{\Omega_c^{(i) \ast}}{2} \langle \sigma^{(i)}_{gd}\rangle + \frac{\Omega_p^{(i)}}{2} \langle \sigma^{(i)}_{de} \rangle \nonumber \\ 
\langle\dot{\sigma}^{(i)}_{eg} \rangle &=& i (\Delta_g - \Delta_e) \langle \sigma^{(i)}_{eg}\rangle + \frac{\Omega_c^{(i) }}{2} \langle \sigma^{(i)}_{dg}\rangle + \frac{\Omega_p^{(i)\ast}}{2} \langle \sigma^{(i)}_{ed} \rangle \nonumber \\
\langle\dot{\sigma}^{(i)}_{4}\rangle &=& -\Gamma_4 \langle\sigma^{(i)}_{1}\rangle  -\Gamma_4 \langle\sigma^{(i)}_{4}\rangle - \frac{\Omega_p^{(i) \ast}}{2} \langle \sigma^{(i)}_{gd} \rangle - \frac{\Omega_p^{(i)}}{2} \langle \sigma^{(i)}_{dg} \rangle \nonumber \\
&& - \Omega_c^{(i) \ast} \langle \sigma^{(i)}_{ed} \rangle - \Omega_c^{(i)} \langle \sigma^{(i)}_{de} \rangle -\Gamma_4 \nonumber \\
\langle\dot{\sigma}^{(i)}_{gd}\rangle &=& \frac{\Omega_p^{(i)}}{2}  \langle\sigma^{(i)}_{1}\rangle -  \frac{\Omega_c^{(i)}}{2} \langle\sigma^{(i)}_{ge}\rangle + (i \Delta_g - \frac{\tilde{\Gamma}}{2}) \langle\sigma^{(i)}_{gd}\rangle \nonumber \\
\langle\dot{\sigma}^{(i)}_{dg}\rangle &=& \frac{\Omega_p^{(i)\ast}}{2}  \langle\sigma^{(i)}_{1}\rangle -  \frac{\Omega_c^{(i)\ast}}{2} \langle\sigma^{(i)}_{eg}\rangle + (-i \Delta_g - \frac{\tilde{\Gamma}}{2}) \langle\sigma^{(i)}_{dg}\rangle \nonumber \\
\langle\dot{\sigma}^{(i)}_{ed}\rangle &=& -\frac{\Omega_p^{(i)}}{2}  \langle\sigma^{(i)}_{eg}\rangle +  \frac{\Omega_c^{(i)}}{2} \langle\sigma^{(i)}_{4}\rangle + (-i \Delta_e - \frac{\tilde{\Gamma}}{2}) \langle\sigma^{(i)}_{ed}\rangle \nonumber \\
\langle\dot{\sigma}^{(i)}_{de}\rangle &=& -\frac{\Omega_p^{(i)\ast}}{2}  \langle\sigma^{(i)}_{ge}\rangle +  \frac{\Omega_c^{(i)\ast}}{2} \langle\sigma^{(i)}_{4}\rangle + (i \Delta_e - \frac{\tilde{\Gamma}}{2}) \langle\sigma^{(i)}_{de}\rangle \nonumber \\
\end{eqnarray}
where $\sigma_1 \equiv \sigma_{dd}-\sigma_{gg}$, $\sigma_4 \equiv \sigma_{dd}-\sigma_{ee}$, $\Gamma_1 \equiv \frac{1}{3} (2\gamma_p+2\Gamma_p+\gamma_c+\Gamma_c)$, $\Gamma_4 \equiv \frac{1}{3} (\gamma_p+\Gamma_p+2\gamma_c+2\Gamma_c)$, and $\tilde{\Gamma}\equiv \gamma_p + \Gamma_p + \gamma_c + \Gamma_c$.  The above equations are linear, so they can be written in a compact matrix form 
\begin{equation}
\langle \dot{\vec{\sigma}}^{(i)} \rangle = \mathbf{M}^{(i)} \langle \vec{\sigma}^{(i)} \rangle +\vec{v}~.
\end{equation}
The bare steady state can be obtained by setting $\langle \dot{\vec{\sigma}}^{(i)} \rangle_\textrm{DS}=0 $, i.e., 
\begin{equation}
\langle\vec{\sigma}^{(i)}\rangle_\textrm{DS} = \frac{-1}{\mathbf{M}^{(i)} } \vec{v}~.
\end{equation}

\section{Combined master equation \label{app:combine}}

After applying the procedures and approximations in the main text, the master equation $\dot{\hat{O}}=\mathcal{L}_b(\hat{O})+\mathcal{L}^{(i)}_a(\hat{O})$ can be written in a time-local form.  The expectation value of any system operator in the form of $\hat{O}_b$ or $\vec{O}\equiv \hat{O}_b \otimes \vec{\sigma}^{(i)}$ varies as
\begin{eqnarray}\label{eq:mean_Ob}
\langle \dot{\hat{O}}_b \rangle &=&i \nu \langle[\hat{b}^\dag \hat{b}, \hat{O}_b]\rangle +\frac{|\mu_p|^2+|\mu_c|^2}{2}\langle \mathcal{D}[\hat{b}+\hat{b}^\dag](\hat{O}_b)\rangle \nonumber \\ &&- i\mu_p \langle \mathcal{A}_p^{(1)\dag} [\hat{b}+\hat{b}^\dag,\hat{O}_b] \rangle - i\mu_p^\ast \langle [\hat{b}+\hat{b}^\dag,\hat{O}_b] \mathcal{A}_p^{(1)}\rangle \nonumber \\
\end{eqnarray}
\begin{eqnarray}
\langle\dot{\vec{O}}\rangle &=& i \nu \langle[\hat{b}^\dag \hat{b}, \vec{O}]\rangle +\frac{|\mu_p|^2+|\mu_c|^2}{2}\langle \mathcal{D}[\hat{b}+\hat{b}^\dag](\vec{O})\rangle \nonumber \\
&&- i\mu_p \langle \mathcal{A}_p^{(1)\dag} [\hat{b}+\hat{b}^\dag,\vec{O}] \rangle - i\mu_p^\ast \langle [\hat{b}+\hat{b}^\dag,\vec{O}] \mathcal{A}_p^{(1)}\rangle \nonumber \\
&&+\mathbf{M}^{(i)}\langle \vec{O}\rangle + \langle\hat{O}_b\rangle \vec{v} - \sqrt{\gamma_p}  \Big( e^{i\omega_{p0} \frac{x_i}{c}} \mathbf{M}_{pd}\langle\mathcal{A}_p^{(i+1)\dag} \vec{O} \rangle\nonumber \\
&& -e^{-i\omega_{p0} \frac{x_i}{c}} \mathbf{M}_p \langle \vec{O} \mathcal{A}_p^{(i+1)}\rangle \Big) \nonumber \\
&&- \sqrt{\gamma_c} \Big( e^{-i\omega_{c0} \frac{x_i}{c}} \mathbf{M}_{cd} \langle\mathcal{A}_c^{(i)\dag} \vec{O}\rangle -e^{i\omega_{c0} \frac{x_i}{c}} \mathbf{M}_c \langle \vec{O} \mathcal{A}_c^{(i)}\rangle \Big) \nonumber \\
&&+ i\frac{\sqrt{\gamma_c}}{2}\big(\mu_c^\ast e^{-i\omega_{c0} \frac{x_i}{c}}\mathbf{M}_{cd}\langle(e^{i\nu\tau}\hat{b} + e^{-i\nu\tau}\hat{b}^\dag) \vec{O}\rangle \nonumber \\
&&+ \mu_c e^{i\omega_{c0} \frac{x_i}{c}} \mathbf{M}_c\langle \vec{O} (e^{i\nu\tau}\hat{b}+e^{-i\nu\tau}\hat{b}^\dag)\rangle \big)  ~. \label{eq:mean_Ov}
\end{eqnarray}
The transformation matrices are defined as $\mathbf{M}_{pd}\vec{\sigma}\equiv [\sigma_{gd},\vec{\sigma}]$, $\mathbf{M}_{p}\vec{\sigma}\equiv [\sigma_{dg},\vec{\sigma}]$, $\mathbf{M}_{cd}\vec{\sigma}\equiv [\sigma_{ed},\vec{\sigma}]$, and $\mathbf{M}_{c}\vec{\sigma}\equiv [\sigma_{de},\vec{\sigma}]$, i.e.,
\begin{equation}
\mathbf{M}_{pd}=
\left(
\begin{array}{cccccccc}
 0 & 0 & 0 & 0 & 2 & 0 & 0 & 0 \\
 0 & 0 & 0 & 0 & 0 & 0 & 0 & 0 \\
 0 & 0 & 0 & 0 & 0 & 0 & -1 & 0 \\
 0 & 0 & 0 & 0 & 1 & 0 & 0 & 0 \\
 0 & 0 & 0 & 0 & 0 & 0 & 0 & 0 \\
 -1 & 0 & 0 & 0 & 0 & 0 & 0 & 0 \\
 0 & 0 & 0 & 0 & 0 & 0 & 0 & 0 \\
 0 & 1 & 0 & 0 & 0 & 0 & 0 & 0 \\
\end{array}
\right)
\end{equation}
\begin{equation}
\mathbf{M}_{p}=
\left(
\begin{array}{cccccccc}
 0 & 0 & 0 & 0 & 0 & -2 & 0 & 0 \\
 0 & 0 & 0 & 0 & 0 & 0 & 0 & 1 \\
 0 & 0 & 0 & 0 & 0 & 0 & 0 & 0 \\
 0 & 0 & 0 & 0 & 0 & -1 & 0 & 0 \\
 1 & 0 & 0 & 0 & 0 & 0 & 0 & 0 \\
 0 & 0 & 0 & 0 & 0 & 0 & 0 & 0 \\
 0 & 0 & -1 & 0 & 0 & 0 & 0 & 0 \\
 0 & 0 & 0 & 0 & 0 & 0 & 0 & 0 \\
\end{array}
\right)
\end{equation}
\begin{equation}
\mathbf{M}_{cd}= 
\left(
\begin{array}{cccccccc}
 0 & 0 & 0 & 0 & 0 & 0 & 1 & 0 \\
 0 & 0 & 0 & 0 & -1 & 0 & 0 & 0 \\
 0 & 0 & 0 & 0 & 0 & 0 & 0 & 0 \\
 0 & 0 & 0 & 0 & 0 & 0 & 2 & 0 \\
 0 & 0 & 0 & 0 & 0 & 0 & 0 & 0 \\
 0 & 0 & 1 & 0 & 0 & 0 & 0 & 0 \\
 0 & 0 & 0 & 0 & 0 & 0 & 0 & 0 \\
 0 & 0 & 0 & -1 & 0 & 0 & 0 & 0 \\
\end{array}
\right)
\end{equation}
\begin{equation}
\mathbf{M}_{c}= 
\left(
\begin{array}{cccccccc}
 0 & 0 & 0 & 0 & 0 & 0 & 0 & -1 \\
 0 & 0 & 0 & 0 & 0 & 0 & 0 & 0 \\
 0 & 0 & 0 & 0 & 0 & 1 & 0 & 0 \\
 0 & 0 & 0 & 0 & 0 & 0 & 0 & -2 \\
 0 & -1 & 0 & 0 & 0 & 0 & 0 & 0 \\
 0 & 0 & 0 & 0 & 0 & 0 & 0 & 0 \\
 0 & 0 & 0 & 1 & 0 & 0 & 0 & 0 \\
 0 & 0 & 0 & 0 & 0 & 0 & 0 & 0 \\
\end{array}
\right)~.
\end{equation}

We have assumed that both the input field of Probe, Control, and baths are vacuum, i.e., $\hat{a}_{\pm y}^\textrm{in}|\textrm{vac}\rangle = \hat{r}_y^\textrm{in}|\textrm{vac}\rangle =0$, for $y$ could be $p$ or $c$.  The round-trip traveling time of radiation between the mirror (at $x=0$) and atomic cloud (mean location $x=\bar{x}$) is $\tau \equiv 2 \bar{x}/c$.  The tilde for the advanced operators has been omitted.

\section{Derivation of recurrence relation \label{app:recurrence}}

To derive the recurrence relation in Eqs.~(\ref{eq:Apb2})-(\ref{eq:Acbd2}), we start by using Eq.~(\ref{eq:mean_Ov}) for $\vec{O}=\hat{b}\otimes \vec{\sigma}$ and $\hat{b}^\dag\otimes \vec{\sigma}$.  We apply the steady state approximation $\sigma\approx \langle\textrm{DS}| \sigma|\textrm{DS}\rangle$, and collect only the leading order quantum correction of atomic operators.  Then we get the following relations for the zero frequency components,
\begin{widetext}
\begin{eqnarray}
2 \pi \langle\hat{b}\sigma_{gd}^{(i)}\rangle_0 &=& -\sqrt{\gamma_p} (e^{i\omega_{p0} \frac{x_i}{c}} G^{(i)}_{pd}(-\nu) \langle \hat{b} \mathcal{A}_p^{(i+1)\dag} \rangle_0 - e^{-i\omega_{p0} \frac{x_i}{c}} G^{(i)}_{p}(-\nu) \langle \hat{b} \mathcal{A}_p^{(i+1)} \rangle_0) \nonumber \\
&&-\sqrt{\gamma_c} (e^{-i\omega_{c0} \frac{x_i}{c}} G^{(i)}_{cd}(-\nu) \langle \hat{b} \mathcal{A}_c^{(i)\dag} \rangle_0 - e^{i\omega_{c0} \frac{x_i}{c}} G^{(i)}_{c}(-\nu) \langle \hat{b} \mathcal{A}_c^{(i)} \rangle_0) \nonumber \\
&&+ i\frac{\sqrt{\gamma_c}}{2}\big(e^{-i\nu\tau} \mu_c^\ast e^{-i\omega_{c0} \frac{x_i}{c}} G^{(i)}_{cd}(-\nu) \langle\hat{b}^\dag \hat{b}\rangle + e^{-i\nu\tau} \mu_c e^{i\omega_{c0} \frac{x_i}{c}} G^{(i)}_c(-\nu) \langle\hat{b} \hat{b}^\dag\rangle \big)~, \\
2 \pi \langle\hat{b}^\dag\sigma_{gd}^{(i)}\rangle_0 &=& -\sqrt{\gamma_p} (e^{i\omega_{p0} \frac{x_i}{c}} G^{(i)}_{pd}(\nu) \langle \hat{b}^\dag \mathcal{A}_p^{(i+1)\dag} \rangle_0 - e^{-i\omega_{p0} \frac{x_i}{c}} G^{(i)}_{p}(\nu) \langle \hat{b}^\dag \mathcal{A}_p^{(i+1)} \rangle_0) \nonumber \\
&&-\sqrt{\gamma_c} (e^{-i\omega_{c0} \frac{x_i}{c}} G^{(i)}_{cd}(\nu) \langle \hat{b}^\dag \mathcal{A}_c^{(i)\dag} \rangle_0 - e^{i\omega_{c0} \frac{x_i}{c}} G^{(i)}_{c}(\nu) \langle \hat{b}^\dag \mathcal{A}_c^{(i)} \rangle_0) \nonumber \\
&&+ i\frac{\sqrt{\gamma_c}}{2}\big(e^{i\nu\tau}\mu_c^\ast e^{-i\omega_{c0} \frac{x_i}{c}} G^{(i)}_{cd}(\nu) \langle\hat{b}\hat{b}^\dag\rangle + e^{i\nu\tau}\mu_c e^{i\omega_{c0} \frac{x_i}{c}} G^{(i)}_c(\nu) \langle\hat{b}^\dag  \hat{b}\rangle \big)~, \\
2 \pi \langle\hat{b}\sigma_{ed}^{(i)}\rangle_0 &=& -\sqrt{\gamma_p} (e^{i\omega_{p0} \frac{x_i}{c}} F^{(i)}_{pd}(-\nu) \langle \hat{b} \mathcal{A}_p^{(i+1)\dag} \rangle_0 - e^{-i\omega_{p0} \frac{x_i}{c}} F^{(i)}_{p}(-\nu) \langle \hat{b} \mathcal{A}_p^{(i+1)} \rangle_0) \nonumber \\
&&-\sqrt{\gamma_c} (e^{-i\omega_{c0} \frac{x_i}{c}} F^{(i)}_{cd}(-\nu) \langle \hat{b}\mathcal{A}_c^{(i)\dag} \rangle_0 - e^{i\omega_{c0} \frac{x_i}{c}} F^{(i)}_{c}(-\nu) \langle \hat{b} \mathcal{A}_c^{(i)} \rangle_0) \nonumber \\
&&+ i\frac{\sqrt{\gamma_c}}{2}\big(e^{-i\nu\tau}\mu_c^\ast e^{-i\omega_{c0} \frac{x_i}{c}} F^{(i)}_{cd}(-\nu) \langle \hat{b}^\dag \hat{b}\rangle + e^{-i\nu\tau}\mu_c e^{i\omega_{c0} \frac{x_i}{c}} F^{(i)}_c(-\nu) \langle\hat{b} \hat{b}^\dag \rangle \big)~, \\
2 \pi \langle \hat{b}^\dag \sigma_{ed}^{(i)} \rangle_0 &=& -\sqrt{\gamma_p} (e^{i\omega_{p0} \frac{x_i}{c}} F^{(i)}_{pd}(\nu) \langle \hat{b}^\dag \mathcal{A}_p^{(i+1)\dag} \rangle_0 - e^{-i\omega_{p0} \frac{x_i}{c}} F^{(i)}_{p}(\nu) \langle \hat{b}^\dag \mathcal{A}_p^{(i+1)} \rangle_0) \nonumber \\
&&-\sqrt{\gamma_c} (e^{-i\omega_{c0} \frac{x_i}{c}} F^{(i)}_{cd}(\nu) \langle \hat{b}^\dag \mathcal{A}_c^{(i)\dag} \rangle_0 - e^{i\omega_{c0} \frac{x_i}{c}} F^{(i)}_{c}(\nu) \langle \hat{b}^\dag \mathcal{A}_c^{(i)} \rangle_0) \nonumber \\
&&+ i\frac{\sqrt{\gamma_c}}{2}\big(e^{i\nu\tau}\mu_c^\ast e^{-i\omega_{c0} \frac{x_i}{c}} F^{(i)}_{cd}(\nu) \langle \hat{b} \hat{b}^\dag\rangle + e^{i\nu\tau}\mu_c e^{i\omega_{c0} \frac{x_i}{c}} F^{(i)}_c(\nu) \langle\hat{b}^\dag \hat{b}\rangle \big)~,
\end{eqnarray}
\end{widetext}
where
\begin{eqnarray} \label{eq:G}
G^{(i)}_y(\omega) &\equiv& 2\pi \hat{u}_5\cdot\frac{-1}{i\omega+\mathbf{M}^{(i)}}\mathbf{M}_y\langle \textrm{DS}|^{(i)}\vec{\sigma}^{(i)}|\textrm{DS}\rangle^{(i)} \\
F^{(i)}_y(\omega) &\equiv& 2\pi \hat{u}_7\cdot\frac{-1}{i\omega+\mathbf{M}^{(i)}}\mathbf{M}_y\langle\textrm{DS}|^{(i)} \vec{\sigma}^{(i)}|\textrm{DS}\rangle^{(i)}~,~~ \label{eq:F}
\end{eqnarray}
for $y=p, pd, c,$ or $cd$.
The projection vectors are defined as $\hat{u}_5 \equiv (0~0~0~0~1~0~0~0)$ and $\hat{u}_7 \equiv (0~0~0~0~0~0~1~0)$, such that $\hat{u}_5\cdot \vec{\sigma}=\sigma_{gd}$ and $\hat{u}_7\cdot \vec{\sigma}=\sigma_{ed}$.  

Evaluating Eqs.~(\ref{eq:G}) and (\ref{eq:F}), we find that $G_{pd}=G_{cd}=F_{pd}=F_{cd}=0$, and
\begin{eqnarray}
G_p^{(i)}(\omega) &=&|\tilde{\alpha}_c|^2 \frac{2\pi}{\gamma_p}J(\omega) \\
G_c^{(i)}(\omega) &=& -e^{-i(\omega_{p0}+\omega_{c0})\frac{x_i}{c}} \tilde{\alpha}_p \tilde{\alpha}_c^\ast \frac{2\pi}{\sqrt{\gamma_p\gamma_c}} J(\omega)  \\
F_p^{(i)}(\omega) &=& -e^{i(\omega_{p0}+\omega_{c0})\frac{x_i}{c}} \tilde{\alpha}_p^\ast \tilde{\alpha}_c \frac{2\pi}{\sqrt{\gamma_p\gamma_c}} J(\omega)  \\
F_c^{(i)}(\omega) &=&|\tilde{\alpha}_p|^2 \frac{2\pi}{\gamma_c}J(\omega)~,
\end{eqnarray}
where $J(\omega)$ is given in Eq.~(\ref{eq:J_exp}).  The recurrence relation Eqs.~(\ref{eq:Apb2}) and (\ref{eq:Acb2}) can be obtained by using the definitions, $ \mathcal{A}_p^{(i)}  -  \mathcal{A}_p^{(i+1)}  = \sqrt{\gamma_p} e^{i \omega_{p0}\frac{x_i}{c}} \sigma_{gd}^{(i)}$ and $ \mathcal{A}_c^{(i+1)}  -  \mathcal{A}_c^{(i)} = \sqrt{\gamma_c} e^{-i \omega_{c0}\frac{x_i}{c}} \sigma_{ed}^{(i)} $.

\section{Solution of recurrence relation \label{app:solution}}

To obtain the value of $\langle \hat{b}\mathcal{A}_p^{(1)} \rangle_0$ from the recurrence relation Eqs.~(\ref{eq:Apb2}) and (\ref{eq:Acb2}), we consider the sum of these equations, that is the relation in Eq.~(\ref{eq:transfer_red}).  This relation implies that the properties at the ends of the atomic cloud are related as
\begin{eqnarray}\label{eq:recurrence_-}
\langle \hat{b} \mathcal{A}_p^{(1)} \rangle_0 - \langle \hat{b} \mathcal{A}_c^{(1)} \rangle_0 &=&\langle \hat{b} \mathcal{A}_p^{(i)} \rangle_0 - \langle \hat{b} \mathcal{A}_c^{(i)} \rangle_0 \nonumber \\
&=&\langle \hat{b} \mathcal{A}_p^{(N_p+1)} \rangle_0 - \langle \hat{b} \mathcal{A}_c^{(N_p+1)} \rangle_0 ~.
\end{eqnarray}
We also consider the difference of Eqs.~(\ref{eq:Apb2}) and (\ref{eq:Acb2}), which gives the following relation
\begin{eqnarray}\label{eq:recurrence_+}
&&\langle \hat{b} \mathcal{A}_p^{(i+1)} \rangle_0 + \langle \hat{b} \mathcal{A}_c^{(i+1)} \rangle_0 \nonumber \\
&\approx &\langle \hat{b} \mathcal{A}_p^{(i)} \rangle_0 + \langle \hat{b} \mathcal{A}_c^{(i)} \rangle_0   - 2 |\tilde{\alpha}|^2 J(-\nu)(\langle \hat{b} \mathcal{A}_p^{(i)} \rangle_0 - \langle \hat{b} \mathcal{A}_c^{(i)} \rangle_0) \nonumber \\
&&+ i e^{-i\nu \tau} \mu_c |\tilde{\alpha}|^2 J(-\nu) \langle \hat{b}\hat{b}^\dag\rangle~.~~
\end{eqnarray}
We have collected only the leading order of the small factor $|\tilde{\alpha}|^2 |J(\omega)| \ll 1$, which is valid because each atom is weakly interacting with quantum radiation.  By repeating Eq.~(\ref{eq:recurrence_+}) from $i=1$ to $i=N_p$, and using Eq.~(\ref{eq:recurrence_-}), we get another relation for the properties at the ends of the atomic cloud
\begin{eqnarray}\label{eq:recurrence_+end}
&&\langle \hat{b} \mathcal{A}_p^{(N_p+1)} \rangle_0 + \langle \hat{b} \mathcal{A}_c^{(N_p+1)} \rangle_0 \nonumber \\
&=& (1- 2 N_p |\tilde{\alpha}|^2 J(-\nu))\langle \hat{b} \mathcal{A}_p^{(1)} \rangle_0 \nonumber\\
&&+ (1+2 N_p |\tilde{\alpha}|^2 J(-\nu)) \langle \hat{b} \mathcal{A}_c^{(1)} \rangle_0   \nonumber \\
&&+ i e^{-i\nu \tau} \mu_c N_p |\tilde{\alpha}|^2 J(-\nu) \langle \hat{b}\hat{b}^\dag\rangle~.
\end{eqnarray}
Combining Eqs.~(\ref{eq:recurrence_-}) and (\ref{eq:recurrence_+end}) to eliminate $\langle \hat{b} \mathcal{A}_c^{(N_p+1)} \rangle_0$, and using the definition $\mathcal{A}_p^{(N_p+1)}=\mathcal{A}_c^{(1)}=0$, we obtain the solution in Eq.~(\ref{eq:bAp}).

Eq.~(\ref{eq:bdAp}) can be obtained through similar procedures.

\bibliographystyle{apsrev4-1}
\pagestyle{plain}
\bibliography{r_cool}

\end{document}